\documentclass[aps,prl,twocolumn,twoside,notitlepage,superscriptaddress, preprintnumbers,showpacs,
showkeys,byrevtex,floatfix,amsmath,amssymb]
{revtex4-1}
\usepackage{array,amscd,amsmath,amssymb,graphicx,mathtools, multirow,placeins,slashed,verbatim,subfig,tensor,dsfont}
\usepackage[bookmarksnumbered,bookmarksopen=true,bookmarksopenlevel=1,pdfstartview=FitH,breaklinks=true]{hyperref}
\usepackage[all]{hypcap}
\usepackage{rotating}
 
\newcolumntype{M}[1]{>{$}{#1}<{$}}

\newcommand{\sst}[1]{{\scriptscriptstyle #1}}

\def\0{{\sst{(0)}}}
\def\1{{\sst{(1)}}}
\def\2{{\sst{(2)}}}
\def\3{{\sst{(3)}}}
\def\4{{\sst{(4)}}}
\def\5{{\sst{(5)}}}
\def\6{{\sst{(6)}}}
\def\7{{\sst{(7)}}}

\newcommand{\be}{\begin{equation}}
\newcommand{\ee}{\end{equation}}
\def\ba{\begin{array}}
\def\ea{\end{array}}

\newcommand{\bea}{\begin{eqnarray}}
\newcommand{\eea}{\end{eqnarray}}

\DeclareMathOperator{\tr}{tr}

\DeclareMathOperator{\OSp}{OSp}

\begin{document}

\title{Gravity as Gauge Theory Squared: A Ghost Story}

\author{A. Anastasiou}
\email[]{alexandros.anastasiou@su.se}
\affiliation{Nordita, KTH Royal Institute of Technology and Stockholm University, Roslagstullsbacken 23, 10691 Stockholm, Sweden}
\author{L. Borsten}
\email[]{leron@stp.dias.ie}
\affiliation{School of Theoretical Physics, Dublin Institute for Advanced Studies,
10 Burlington Road, Dublin 4, Ireland}
\author{M. J. Duff}
\email[]{m.duff@imperial.ac.uk}
\affiliation{Theoretical Physics, Blackett Laboratory, Imperial College London,
London SW7 2AZ, United Kingdom}
\affiliation{Mathematical Institute, University of Oxford, Andrew Wiles Building, Woodstock Road, Radcliffe Observatory Quarter,
Oxford, OX2 6GG, United Kingdom}
\author{S. Nagy}
\email[]{silvia.nagy1@nottingham.ac.uk}
\affiliation{Centre for Astronomy \& Particle Theory,
University Park,
Nottingham,
NG7 2RD,
United Kingdom}
\author{M. Zoccali}
\email[]{m.zoccali14@imperial.ac.uk}
\affiliation{Theoretical Physics, Blackett Laboratory, Imperial College London,
London SW7 2AZ, United Kingdom}\date{\today}

\begin{abstract}

 The Becchi-Rouet-Stora-Tyutin (BRST) transformations and equations of motion of  a gravity--two-form--dilaton system are derived from the product of two  Yang-Mills theories in a BRST covariant form, to linear approximation.   The inclusion of ghost fields facilitates the separation of the graviton and dilaton. The gravitational gauge fixing term is uniquely determined by those of the Yang-Mills factors which can be freely chosen.  Moreover, the resulting gravity--two-form--dilaton Lagrangian is anti-BRST invariant and the BRST and anti-BRST charges anti commute as a direct consequence of the formalism. 

\end{abstract}


\preprint{DIAS-STP-18-10, Imperial-TP-2018-MJD-01}

\maketitle

\paragraph{Introduction.---}Is it possible that  gravity can be reformulated as  the ``product'' of two  gauge theories?  Despite the obvious and significant differences of gauge and gravity theories, such a  picture has indeed emerged over recent years, particularly  in the context  of the Bern-Carrasco-Johansson  construction of gravity scattering amplitudes as the ``double copy'' of Yang-Mills amplitudes \cite{Bern:2008qj, Bern:2010ue, Bern:2010yg}. This framework is conceptually suggestive and powerful, as exemplified by the remarkable five-loop  result in $\mathcal{N}=8$ supergravity \cite{Bern:2018jmv}. However, we need not restrict ourselves to amplitudes; there are now complementary formulations of the ``gravity $=$ gauge $\times$ gauge'' paradigm at the level of classical solutions and the spacetime fields themselves \cite{Monteiro:2011pc, BjerrumBohr:2012mg, Monteiro:2013rya, Borsten:2013bp, Monteiro:2014cda, Borsten:2015pla, Luna:2015paa, Cardoso:2016ngt,  Luna:2016due, Luna:2016hge,Anastasiou:2016csv, Cardoso:2016amd, Anastasiou:2017nsz, Borsten:2017jpt, Anastasiou:2017taf, Luna:2017dtq, Carrillo-Gonzalez:2017iyj,  Shen:2018ebu, Li:2018qap}, as well as  strongly coupled  theories with no Lagrangian description \cite{Ferrara:2018iko}.  Such approaches provide a different way of posing the question in what precise sense gravity is  the square of gauge. 

The product of two pure Yang-Mills gauge theories typically yields  gravity coupled to a dilaton and a Kalb-Ramond two-form \footnote{A theory sometimes referred to as $\mathcal{N}=0$ supergravity.}. Consequently,  a thorny issue is the separation of the graviton and dilaton \cite{LopesCardoso:2018xes}.   In the present contribution, we reconsider the graviton--two-form--dilaton system using the field-theoretic realization of {gravity $=$ gauge $\times$ gauge}  introduced in Refs. \cite{Borsten:2013bp, Anastasiou:2014qba}.    
The field-theoretic product of two gauge potentials, $A_\mu$ and  $\tilde{A}_\nu$,   is given by
\begin{equation}\label{product}
[A_\mu \circ \tilde{A}_\nu](x) := [A_\mu^a \cdot \Phi_{a\tilde{a}} \cdot \tilde{A}_\nu^{\tilde{a}}](x).
\end{equation}
Here, $\cdot$ denotes a convolution and $\Phi_{a\tilde{a}}$ a spectator scalar field valued in the bi-adjoint of the Left and Right gauge groups, $G$ and $\tilde{G}$, of   $A_\mu$ and  $\tilde{A}_\nu$, respectively.
The spectator field is closely related, as a convolutive inverse \cite{LopesCardoso:2018xes}, to   the $\phi^3$ theory appearing in the context of  scattering amplitudes \cite{Hodges:2011wm, Cachazo:2013iea, Naculich:2014naa, Anastasiou:2016csv} and  classical  solutions \cite{Monteiro:2011pc, Monteiro:2014cda, Luna:2015paa, Cardoso:2016ngt, Luna:2016due, Cardoso:2016amd, Luna:2016hge, White:2016jzc, Luna:2017dtq, Bahjat-Abbas:2017htu}. Its role here is to allow for arbitrary and independent $G$ and $\tilde{G}$. Using \eqref{product} it was shown that the   local symmetries of the gravity theory (general coordinate, local supersymmetry and $p$-form gauge transformations) follow  from those of the  Yang-Mills factors \cite{Anastasiou:2014qba}, to linear approximation. Our present  aim  is to provide  an explicit map between the  \textit{equations of motion} and  \textit{gauge-fixing choices} of the gauge theory factors and those of the corresponding gravity theory.
Regarding the latter, we seek a formulation of the {gravity $=$ gauge $\times$ gauge} dictionary which is general enough to derive equations of motion of gravity from Yang-Mills {without having to restrict to a specific gauge}, while  disentangling the graviton, dilaton, and two-form fields. As we shall demonstrate, these ambitions  may be realised  by adopting a  Becchi-Rouet-Stora-Tyutin (BRST) \cite{Becchi:1975nq, Tyutin:1975qk, Fradkin:1975cq, Batalin:1977pb, Kugo:1979gm, Henneaux:1992ig, Gomis:1994he} formalism and paying due diligence to  boundary conditions.


When considering  general classes of gauge-fixing choices, one is quite naturally led to consider a BRST  approach, even when discussing purely classical aspects of the field theories.  Indeed, the inclusion of ghosts in the context of Yang-Mills squared was long-ago introduced by Siegel \cite{Siegel:1988qu, Siegel:1995px}. Here, we shall construct the physical and ghost fields of the graviton--two-form--dilaton system as the product of Yang-Mills  fields using \eqref{product}.  Then, by exploiting the properties of the  product \eqref{product},  we are able to \emph{derive} the BRST variation and equations of motion  of the gravity theory from those of the Yang-Mills factors alone.  

This approach requires two further ingredients. The  first  concerns the domain of validity of the ``derivative rule" of the convolution, which  is \emph{not} Leibniz, but rather satisfies
\begin{equation}\label{derivative_rule}
\partial_\mu (f\cdot g) = (\partial_\mu f) \cdot g = f \cdot (\partial_\mu g).
\end{equation}
This property is key to recovering both the BRST variations and equations of motion. However, Eq. \eqref{derivative_rule} only holds for sufficiently well-behaved functions. 
Following the treatment given in Ref. \cite{LopesCardoso:2018xes}, one can suitably restrict to a well-defined domain (by, for example, introducing punctures) and encode the associated boundary conditions through the   introduction of effective sources, generically denoted here by $j$. Second, the convolution dictionary realized through Eq. \eqref{product} requires the inclusion of the  nonlocal Green's  operator $ G \cdot j \equiv \Box^{-1}j$ \cite{LopesCardoso:2018xes}, which  commutes with  the d'Alembertian,  
\begin{equation}
\Box^{-1} \Box = \Box  \Box^{-1} = \text{Id},
\end{equation}
when acting on fields not in the kernel of $\Box$. These ingredients combine with the inclusion of ghosts to single out a unique gravity $=$ gauge $\times$ gauge field map, as we shall demonstrate in the following.

\paragraph{The gauge theories.---}Consider two copies of linearized Yang-Mills theory with the BRST Lagrangian 
\be
\mathcal{L}=\tr \left( -\tfrac{1}{4}F^{\mu\nu} F_{\mu\nu} - b G(A, b) - \bar{c} \int d^4y \tfrac{\delta G}{\delta A^\mu}\partial^\mu c\right),
\ee
where $F^a_{\mu\nu} = 2\partial_{[\mu} A^a_{\nu]}$ \footnote{The linearisation  makes the non-Abelian part of the symmetry global.} and $c$, $b$, and $G$ are the ghost, Lautrup-Nakanishi Lagrange multiplier and gauge fixing function, respectively.  The gauge group indices will be suppressed in the following. 
While not necessary,   we work within the one-parameter $\xi$ family of  general linear covariant gauges, the most general class maintaining manifest Lorentz covariance. Then, after elimination of the Lautrup-Nakanishi auxiliary field, $b$, we have,
\begin{equation}\label{YM_lagrangian}
\mathcal{L}_{A} = \tr \left( -\frac{1}{4}F^{\mu\nu} F_{\mu\nu} + \frac{1}{2\xi}(\partial^\mu A_\mu)^2 - \bar{c}\Box c \right).
\end{equation}
 As usual with BRST, the choice of gauge-fixing (here a Gaussian average over Lorenz gauge) is implemented via the insertion of a delta functional in the path integral, compensated by a Jacobian factor (lifted to the action via the ghosts, $c$), ensuring that physical observables are unaffected by the gauge choice (i.e., the $\xi$ dependence drops out). Variation of the above leads to the equations of motion,
\begin{equation}\label{YM_eom}
\Box A_\mu - \tfrac{\xi+1}{\xi}\partial_\mu \partial A = j_\mu (A), \quad
\Box c^{\alpha} = j^{\alpha} (c)
\end{equation}
where, for simplicity, we have added the sources directly into the equations \footnote{Note, it is possible to add them to the Lagrangian while preserving the BRST invariance.} and  introduced the  $\OSp(2)$ ghost-antighost doublet $c^{\alpha}$,  where $c^{1}=c, c^{2}=\bar{c}$ \cite{Hull:1990hc}. As indicated previously, the sources ensure that the convolution is well-defined and obeys Eq. \eqref{derivative_rule} by effectively encoding boundary conditions.
The theory is invariant under the BRST transformations
\begin{equation}\label{YM_BRST}
Q A_\mu = \partial_\mu c, \qquad Qc = 0, \qquad Q \bar{c} = \frac{1}{\xi}\partial^\mu A_\mu.
\end{equation}
Thus, squaring two pure Yang-Mills theories now involves \textit{all possible} products between the sets of fields $(A_\mu, c^{\alpha})$, $\Phi$ and $(\tilde{A}_\nu, \tilde{c}^{\beta})$. As first noted in Refs. \cite{Siegel:1988qu, Siegel:1995px},  the degrees of freedom,  ghost number and parity inherited by the products are very suggestive that squaring two BRST-covariant Yang-Mills theories results in the states, physical as well as first- and second-level ghosts, of a graviton, two-form and dilaton. As an interesting example, the ghost-antighost triplet in the $\underline{2} \otimes \underline{2}$ of $\OSp(2)$, given by  $c^{(\alpha} \circ \tilde{c}^{\beta)}$, corresponds to the {three} second-level ghosts of the two-form theory, while the singlet, $c^\alpha\circ \tilde{c}_\alpha$, yields an auxiliary  degree of freedom, which, as we shall see, contributes to both the graviton \eqref{grav} and the dilaton \eqref{dila}.

\paragraph{The gravity theory.---}First, we give the BRST Lagrangians for the graviton $h_{\mu\nu}$  and dilaton $\varphi$ (defined around Minkowski) and the two-form, $B_{\mu\nu}$. The former, in Einstein frame, reads
\begin{equation}\label{Lh}
\begin{split}
\mathcal{L}_{h, \varphi} = &-\frac{1}{4}h^{\mu\nu}E_{\mu\nu} + \frac{1}{2\xi_{(h)}}\left( \partial^\nu h_{\mu\nu} - \frac{1}{2}\partial_\mu h \right)^2\\
& - \frac{1}{4} (\partial\varphi)^2- \bar{c}^\mu \Box c_{\mu}\
\end{split}
\end{equation}
where $E_{\mu\nu}$ is the linearized Einstein tensor and we  average over de Donder gauge fixings, controlled by $\xi_{(h)}$. For the latter,
\begin{equation}\label{LB}
\begin{split}
\mathcal{L}_{B} &= -\frac{1}{24}H^{\mu\nu\rho}H_{\mu\nu\rho}  \\
&+ \frac{1}{2\xi_{(B)}}\left( \partial_\mu B^{\mu\nu} + \partial^\nu \eta \right)^2 - \bar{d}_\nu \Box d^\nu  \\
&+ \frac{\xi_{(d)} - m_{(d)}}{\xi_{(d)}} \bar{d}_\mu \partial^\mu \partial^\nu d_\nu + m_{(d)}\bar{d}\Box d, 
\end{split}
\end{equation}
where one notices the first-level ghosts for the two-form gauge invariance $(d_\mu, \bar{d}_\mu)$ at ghost number $(1, -1)$, as well as the second-level bosonic ghosts, $(d, \bar{d}, \eta)$ at ghost number $(2, {-}2, 0)$, respectively. The gauge fixing of the two-form invariance is controlled by $\xi_{(B)}$, while that of the residual ghost invariance by $\xi_{(d)}$. We consider an extra parameter, $m_{(d)}$, whose relevance is ultimately related to anti-BRST invariance, as discussed below. The equations of motion are given by
\begin{subequations}\label{eom}
\begin{align} \label{eom1}
&\Box h_{\mu\nu} - \tfrac{\xi_{(h)} + 2}{\xi_{(h)}}\left(2 \partial^\rho \partial_{(\mu}h_{\nu)\rho} - \partial_\mu \partial_\nu h \right)= j_{\mu\nu}(h),  \\ \label{eom2}
&\Box B_{\mu\nu} +\tfrac{\xi_{(B)} + 2}{\xi_{(B)}}2\partial^\rho \partial_{[\mu} B_{\nu]\rho} = j_{\mu\nu}(B), \\ \label{eom3}
&\Box \varphi = j(\varphi), 
\end{align}
\end{subequations}
complemented by those for the ghosts, which we omit for brevity.
%
%
The Lagrangians, \eqref{Lh} and \eqref{LB}, are invariant under the BRST transformations
\begin{equation}\label{graviton_BRST}
\begin{array}{llllllll}
Q h_{\mu\nu} &=& 2\partial_{(\mu} c_{\nu)} & Q\varphi &=& 0\\ [5pt]
Q c_\mu &=& 0& Q \bar{c}_\mu&=& \frac{\left( \partial^\nu h_{\mu\nu} - \frac{1}{2}\partial_\mu h \right)}{\xi_{(h)}}\\[5pt]
Q B_{\mu\nu} &=& 2\partial_{[\mu} d_{\nu]}& Q d_\mu & =& \partial_\mu d \\[5pt]
Q \bar{d}_\mu &=& \frac{\left( \partial^\nu B_{\nu\mu} + \partial_\mu \eta \right) }{\xi_{(B)}}&Q d & =& 0\\[5pt]
Q \bar{d} &=& \frac{1}{\xi_{(d)}}\partial^\mu \bar{d}_\mu & Q\eta & =& \frac{m_{(d)}}{\xi_{(d)}}\partial^\mu d_{\mu}. 
\end{array}
\ee
Note,  our choice of Einstein frame implies the BRST invariance of the dilaton, $Q\varphi = 0$. We can go to string frame, where $Q\varphi = \partial^\mu c_\mu$, via a field redefinition, but at linear order such redefinitions are trivial sums.


\paragraph{The {gravity $=$ gauge $\times$ gauge}  dictionary.---}We are now in the position   to construct a {gravity $=$ gauge $\times$ gauge}  dictionary such that the gravitational equations \eqref{eom} and BRST variations \eqref{graviton_BRST} are derived from those of the  Yang-Mills factors in Eqs. \eqref{YM_eom} and \eqref{YM_BRST}.

We begin with the simplest case of the dilaton to clarify the main points. Crucially,  the dilaton dictionary must necessarily be able to reproduce Eq. \eqref{eom3}, without appealing to a convenient choice of  $\xi$. For example, the naive  dictionary  $\varphi = A^\rho \circ \tilde{A}_\rho + \alpha\ c^\alpha \circ \tilde{c}_\alpha$ will not suffice, as applying  the d'Alembertian  reproduces \eqref{eom3} only for a specific  value of $\xi$. The most general ansatz compatible with the required tensor structure, mass dimension, and ghost number is given by 
\begin{equation}\label{dictionary_dilaton}
\varphi = A^\rho \circ \tilde{A}_\rho + \alpha_1 c^\alpha \circ \tilde{c}_\alpha + \frac{\alpha_2}{\Box}\partial A \circ \partial \tilde{A}
\end{equation}
where, crucially, we rely on  $\Box^{-1}$ to allow for the $\partial A \circ \partial \tilde{A}$ term. Note, while it has no bearing on Eq. \eqref{dictionary_dilaton}  we will treat the left and right factors democratically, so that the dictionary is invariant under the interchange of the left and right Yang-Mills theories.  Similarly, the most general ansatz up to  $\partial^\rho j_\rho = -\Box \partial A/\xi$, the analog of current conservation after BRST quantization, for the dilaton effective source in Eq. \eqref{eom3} is given by 
\begin{equation}
j(\varphi) = \frac{\alpha_3}{\Box}j^\rho \circ \tilde{j}_\rho + \frac{\alpha_4}{\Box}j^\alpha \circ \tilde{j}_\alpha.
\end{equation}
On applying $\Box$ to Eq. \eqref{dictionary_dilaton} we obtain three distinct terms 
\begin{subequations}
\begin{align}
\begin{split}
\Box(A^\rho \circ \tilde{A}_\rho) 
&= \left(1-\tfrac{1}{\xi^2}\right) \partial A \circ \partial \tilde{A} + \frac{1}{\Box} j^\rho \circ \tilde{j}_{\rho}
\end{split}
\\
\Box(c^\alpha \circ \tilde{c}_\alpha) &= \frac{1}{\Box}j^\alpha(c) \circ \tilde{j}_\alpha(c) \\
\Box(\Box^{-1}\partial A \circ \partial \tilde{A}) &= \partial A \circ \partial \tilde{A}
\end{align}
\end{subequations}
where we have used   the Yang-Mills equations \eqref{YM_eom}.

The independent  (up to equations of motion) tensor structures   behave much like an orthogonal basis, in that the coefficient of each such term has to vanish separately to satisfy $\Box \varphi = j(\varphi)$.  Imposing the dilaton equation of motion  \eqref{eom3} implies $\alpha_3=1$, $\alpha_4=\alpha_1$ and $\alpha_2 = -1+1/\xi^2$. The final arbitrary parameter is fixed to  $\alpha_1 = 1/\xi$ by demanding that the dictionary reproduce the correct gravitational BRST transformation, $Q\varphi = 0$, given the BRST variations of the underlying Yang-Mills fields \eqref{YM_BRST}, the rule \eqref{derivative_rule} and the fact that the BRST operator anticommutes with Grassmann-valued fields. 

To summarize, the dictionary \eqref{dictionary_dilaton} for the dilaton yields the correct gravitational equations of motion and BRST transformations, given  those of the two  Yang-Mills factors. Furthermore, the weights of each term in the dictionary are uniquely determined in terms of the Yang-Mills gauge-fixing parameter $\xi$. Note that this result crucially relies on the introduction of the $\Box^{-1}$ terms. Since at linear level Eq. (10c) is frame independent, a field redefinition cannot be used to remove the nonlocal terms. 

Applying the same reasoning for graviton and two-form, we can then give the final dictionaries  for all three physical fields: 
\begin{enumerate}
\item The graviton
\begin{equation}
\begin{split}\label{grav}
h_{\mu\nu} =\ &A_{(\mu}\circ \tilde{A}_{\nu)} + a_1 \frac{\partial_\mu \partial_\nu}{\Box} A \circ \tilde{A} \\
&+ a_2 \frac{\partial_\mu \partial_\nu}{\Box} c^\alpha \circ \tilde{c}_\alpha \\
&+ \frac{a_3}{\Box} \left( \partial A \circ \partial_{(\mu}\tilde{A}_{\nu)} + \partial_{(\mu} A_{\nu)} \circ \partial \tilde{A} \right) \\
&+ \eta_{\mu\nu} \left( b_1 A \circ \tilde{A} + b_2 c^\alpha \circ \tilde{c}_\alpha + \frac{b_3}{\Box} \partial A \circ \partial \tilde{A} \right)
\end{split} 
\end{equation}
\item The Kalb-Ramond two-form
\begin{equation}
\begin{split}\label{KB}
B_{\mu\nu} =\ &A_{[\mu} \circ \tilde{A}_{\nu]} \\&- \frac{1}{2\Box}  \left( \partial A \circ \partial_{[\mu} \tilde{A}_{\nu]} - \partial_{[\mu} A_{\nu]} \circ \partial \tilde{A} \right)
\end{split} 
\end{equation}
\item The dilaton
\begin{equation}\label{dila}
\varphi =A^\rho \circ \tilde{A}_\rho + \frac{1}{\xi} c^\alpha \circ \tilde{c}_\alpha + \left(\tfrac{1}{\xi^2}-1\right) \frac{1}{\Box}\partial A \circ \partial \tilde{A}
\end{equation}
\end{enumerate}
where
\begin{subequations}
\begin{alignat}{6}
a_1 &= \frac{1}{1-\xi}, \qquad & b_1 &= \frac{\xi}{(2-D)(\xi-1)},\\
  a_2 &= \frac{1+\xi}{2(1-\xi)},  \qquad & b_2 &= b_1/\xi,\\
   a_3 &= -1/2, \qquad & b_3 &= \left(\tfrac{1}{\xi^2}-1\right) b_1.
\end{alignat}
\end{subequations}
As anticipated, these consistently map the Yang-Mills equations of motion \eqref{YM_eom}  into the gravitational  equations of motion  \eqref{eom}. In addition, they map  the Yang-Mills BRST variations \eqref{YM_BRST} into the gravitational BRST variations \eqref{graviton_BRST}, under the action of $Q$; in the process, we read off the dictionaries for the ghosts:
\begin{subequations}
\begin{align}
\begin{split}
c_\mu &=  \frac{1}{4}\left[c \circ \tilde{A}_\mu + A_\mu \circ \tilde{c} \right.\\  & \left. \phantom{= \tfrac{\xi'+1}{\xi'-1}\left[ \right. }  - \tfrac{\xi+1}{\xi}\tfrac{\partial_\mu}{\Box} \left( c \circ \partial \tilde{A} + \partial A \circ \tilde{c} \right) \right]
\end{split}
\\
\begin{split}
d_{\mu} &= \tfrac{1}{4}\left[ c\circ \tilde{A}_\mu - A_\mu \circ \tilde{c} \right.\\&\left. \phantom{= \tfrac{\xi'+1}{4(\xi'-1)}\left[ \right.} - \tfrac{\xi+1}{\xi}\tfrac{\partial_\mu}{\Box}\left( c\circ \partial \tilde{A} - \partial A \circ \tilde{c} \right)\right] \end{split}
\\
d &= \tfrac{1}{2\xi}c \circ \tilde{c} \\
\eta &= \tfrac{1}{4}\left( c\circ \tilde{\bar{c}} + \bar{c}\circ \tilde{c} \right)
\end{align}
\end{subequations}
The corresponding antighosts are equivalently obtained by either  acting with the Yang-Mills anti-BRST operator, $\overline{Q}$, or by replacing $c\rightarrow \bar{c}$. The bosonic ghost $\eta$ has ghost number zero and maps to itself. Once again, the gravity sources are uniquely built out of $j\circ \tilde{j}$ terms and can be determined directly from Eqs. \eqref{grav} and \eqref{KB}. The inclusion of the (anti)ghost sources allows for independent graviton and dilaton sources (cf.~Sec. 4 of Ref. \cite{LopesCardoso:2018xes}) by reintroducing the independent trace of the graviton source in close analogy to the graviton and dilaton field dictionaries given in and Eqs. \eqref{grav} and \eqref{dila}. Note that in the preceding analysis we have restricted to terms of order $\Box^{-1}$. We can, however, consider the complete $\Box^{-n}$ expansion. The only additional, admissible, and  nontrivial term is at order $\Box^{-2}$ and can only be consistently included for the graviton.  Although it introduces  an additional parameter \footnote{Including the unique $\Box^{-2}$ term in the graviton ansatz implies that its parameter, $a_4$, remains unfixed and $a_1 = \xi^2(1-\xi^2)^{-1}a_4+(1-\xi)^{-1};\ a_2 = \xi(1-\xi^2)^{-1}a_4 + (1+\xi)(2-2\xi)^{-1}$ and $b_1 = (2-D)^{-1}\left[ \xi^2(\xi^2-1)^{-1}a_4 + \xi(\xi-1)^{-1} \right]$, while the rest is left unchanged. Setting $a_4=0$ returns the result in the main body.}, it does  not change the present results.

The linearized Riemann tensor is related to the left and right field strengths by
\begin{gather}
\begin{split}
R_{\mu\nu\rho\sigma} = \frac{1}{2}\left(F_{\mu\nu}\circ\tilde{F}_{\rho\sigma}+F_{\rho\sigma}\circ\tilde{F}_{\mu\nu}\right) \\
+\partial_{[\mu}  \partial_{[\nu} \eta_{\sigma]\rho]}  \left( b_1 A \circ \tilde{A} + b_2 c^\alpha \circ \tilde{c}_\alpha + \frac{b_3}{\Box} \partial A \circ \partial \tilde{A} \right)
\end{split}
\end{gather}
and 
\be
F_{\mu\nu}\circ\tilde{F}_{\rho\sigma} = \partial_{[\mu}Z_{\nu][\rho,\sigma]}
\ee
where
\be
Z_{\mu\nu} = h_{\mu\nu} +B_{\mu\nu} - b_1\eta_{\mu\nu}\varphi. 
\ee


\paragraph{Conclusions.---} Using the field-theoretic product of independent Left and Right gauge theories given in \eqref{product} we have shown that the equations of motion and BRST variations of the graviton, two-form and dilaton follow from those of two pure Yang-Mills gauge potentials, to linear approximation.  The dictionary, with our previously stated assumptions, relating the  gravity fields to  those of the Yang-Mills theories is unique up to order $\Box^{-1}$.   The coefficient of  the unique $\Box^{-2}$ term, $\Box^{-2}\partial_\mu \partial_\nu \partial A\circ \partial \tilde{A}$,  that can be included in the graviton ansatz remains unfixed at linear order and can  be freely set to zero, but may be important at the next to linear order or higher.  

Note, we  chose to perform the calculations using a specific one-parameter family of gauges, sufficient to make the pertinent  features of the construction apparent.  This  yields a one-parameter family of gauges for the graviton and two-form. However, we could have  chosen any other family of gauges to start out with. Turning the handle the conclusions would remain the same, although the details would reflect the new choice.  In this sense, our construction is gauge independent. This can be made manifest  by not specifying the gauge-fixing function, $G(A, b)$, at all and then following the exact same logic. 

A novel feature of the formalism is the gratis  appearance of anti-BRST symmetry. Since the Yang-Mills factors are anti-BRST invariant, they have anti-BRST transformations  that will generate  anti-BRST transformations for the graviton and two-form.  Consequently, we obtain an anti-BRST invariant formulation.    Interestingly,  in    the standard treatment of a free Abelian two-form, the BRST and anti-BRST charges do not anticommute  \cite{BarcelosNeto:1998kj}.   As shown in \cite{Bonora:2007hw}, their anticommutation requires a modification similar to the Curci-Ferrari condition for non-Abelian one-forms. Here,  anticommutation is automatically inherited from the Yang-Mills factors and, hence, we must have in fact landed on the modified two-form theory. Indeed, consistency with the equations of motion and BRST variations requires $m_{(d)}=\xi_{(d)}$, reproducing precisely the formalism of \cite{Bonora:2007hw}, with the two-form ghost  in Lorenz gauge. 

Another  feature of this {gravity $=$ gauge $\times$ gauge}    map is the simplicity with which one is able to control the one-parameter family  of gauge fixing: in addition to the dictionaries themselves being fully fixed in terms of $\xi$, one also has the relation
\begin{equation}
\xi_{(h)} = \xi_{(B)} = 2\xi_{(d)} = \xi
\end{equation}
Then, it is compelling to ask which choices of $\xi$ produce especially tractable cases. For instance, we could choose to simplify the Yang-Mills side with $\xi=-1$, corresponding to the Feynman-'t Hooft gauge. Equations \eqref{YM_eom} become harmonic and the propagator is just 
\be
\tilde{D}^{\mu\nu}_F = \frac{-i\eta^{\mu\nu}}{k^2 + i\varepsilon}.
\ee
 In this gauge, the gravity equations of motion and gauge-fixing functionals lose any dependence on the $\Box^{-1}$ terms. Conversely, we can obtain the same simplification on the gravity side by fixing $\xi = -2$:
\begin{equation*}
\Box h_{\mu\nu} = j_{\mu\nu}(h), \quad \Box B_{\mu\nu} = j_{\mu\nu}(B), \quad \Box \varphi = j(\varphi).
\end{equation*}

 There are a number of further directions one might pursue. Most pressing is the need to go to higher orders in perturbation theory. This also motivates the interesting possibility  of extending the map to the whole field--antifield formalism.  For certain applications it may also be useful to allow for distinct (families) of gaugings in the left and right factors. This would necessarily break invariance of the dictionary under the interchange of the left and right factors and would deform the $\OSp(2)$ symmetry of the ghost$\otimes$antighost sector.  Finally, although we have restricted our attention here to bosonic fields, following Refs.\cite{Siegel:1988qu, Siegel:1995px,Anastasiou:2014qba} it would be straightforward to extend to super Yang-Mills theory and supergravity, particularly when an off-shell superfield formalism is available, in which case the auxiliary fields are retained in both the gauge factors and their product.

\acknowledgments

We are grateful for stimulating conversations with Gabriel L.~Cardoso, Alexander Harrold, Gianluca Inverso, Ricardo Monteiro, Suresh Nampuri, Donal O'Connell and Christopher D.~White. 
We are grateful   to Philip Candelas for hospitality at the Mathematical Institute, University of Oxford. M. J. D.~is grateful to the Leverhulme Trust for an Emeritus Fellowship.
The work of L. B.~is supported by a Schr\"odinger Fellowship. The work of M. J. D.~ is supported by was  supported in part by the STFC under rolling Grant No.~ST/P000762/1. S. N.~is supported by a Leverhulme Research Project Grant. M. Z.~is supported by an EPSRC Ph.D. studentship.


%

\end{document}